\documentclass[12pt,preprint]{aastex}

\newcommand{\kms}{km~s$^{-1}$}
\newcommand{\masyr}{mas~yr$^{-1}$}

\shorttitle{Companion to HIP 115147}
\shortauthors{Makarov, Zacharias \& Hennessy}
\begin{document}

\title{The nearby young visual binary HIP 115147 and its common proper motion companion 
LSPM J2322+7847} 
\author{V. V. Makarov}
\affil{Michelson Science Center, California Technology Institute, 770 S. Wilson Ave.,
MS 100-22, Pasadena, CA 91125}
\email{vvm@caltech.edu}
\author{N. Zacharias, G.S. Hennessy}
\affil{US Naval Observatory, 3450 Massachusetts Ave, NW, Washington, DC 20392-5420}
\author{H. C. Harris, A.K.B. Monet}
\affil{US Naval Observatory, Flagstaff, AZ}

\begin{abstract}
We report a late M-type, common proper motion companion to a nearby young visual binary HIP 115147 (V368 Cep), separated by 963 arcseconds from the primary K0 dwarf. This
optically dim star has been identified as a candidate high proper motion, nearby dwarf
LSPM J2322+7847 by L{\'e}pine in 2005. The wide companion is one of the latest post-T Tauri
low mass stars found within 20 pc. We obtain a trigonometric parallax of $51.6\pm0.8$ mas, in
good agreement with the Hipparcos parallax of the primary star ($50.7\pm0.6$ mas).
Our $BVRI$ photometric data and near-infrared data from 2MASS
are consistent with LSPM J2322+7847 
being brighter by 1 magnitude in $K_s$ than field M dwarfs at $V-K_s=6.66$, which indicates its
pre-main sequence status. We conclude
that the most likely age of the primary HIP 115147 and its 11-arcsecond companion HIP 115147B
is 20-50 Myr. The primary appears to be older than 
its close analog PZ Tel (age 12-20 Myr) and members of the TWA association (7 Myr).
\end{abstract}

\label{firstpage}

\section{Introduction}
In the course of an extensive search for very wide, common proper motion companions
to nearby Hipparcos stars from the NOMAD catalog \citep{zac}, we came across an optically faint
and red star at an angular separation of $963\arcsec$, position angle $141\degr$ from the active and rapidly rotating  G9V star HIP 115147 (BD +78 826, HD 220140, V368 Cep). 
This faint star was subsequently identified with LSPM J2322+7847,
a candidate nearby low-mass dwarf detected by \citet{lep05} from the LSPM-North
catalog \citep{lesh}. The original identification
of this star as one with significant proper motion traces to Luyten (1967)
where it was reported as a magnitude 17 object with $\mu=200$ \masyr\ at a position
angle of $67\degr$ and assigned the name LP12-90. 

In the present paper we present new $BVRI$ photometry of HIP 115147, its known visual companion
HIP 115147B and LSPM J2322+7847, and obtain preliminary
trigonometric parallax astrometry for the latter companion.  Then we discuss
the possible young age and origin of this interesting triple system.

\section{Observations}
Following the identification of LSPM J2322+7847 as a potential widely
separated common proper motion companion to HIP 115147, $BVRI$ photometry
on the Cousins system was obtained for the fainter star on UT 30 August 2005
using the 1.0-meter reflector at the Flagstaff Station.  The 
photometry was calibrated with an instrumental zero point term and a 
first order airmass term. The calibration field was the standard field 
PG2213-006 from \citet{lan}. Additional
photometric observations were subsequently obtained on UT 16 and
17 June 2007 when the individual components of the brighter system
HIP 115147AB (Sep.~11") were also measured.  The photometric results are
presented in Table 1 along with $JHK$ photometry extracted from 2MASS, proper
motions from NOMAD and parallax determinations.  The
estimated uncertainties in the $BVRI$ measures are $\pm0.02$ mag in the case
of LSPM J2322+7847 and $\pm0.03$-0.04 mag for the HIP 115147 components
where the short exposure times introduced additional error from scintillation
and shutter timing.

     Since the August 2005 photometry indicated that LSPM J2322+7847 was
most likely an M-dwarf at approximately the same distance as HIP 115147,
it was added to the trigonometric parallax program at USNO's Flagstaff
Station.  Through June 2007, a total of 66 acceptable CCD observations
have been accumulated on this field, covering an epoch range of 1.65
years.  The same Tek2k CCD, observational procedures, and reduction
algorithms have been employed as summarized in \citet{dah}.  Using
a total of 29 reference stars, the current preliminary solution yields
$\Pi_{\rm rel}=50.72\pm0.73$ mas.  This solution appears to be very robust,
with the separate solutions for parallax in the RA and DEC directions in
very satisfactory agreement ($50.72\pm0.77$ mas versus $50.69\pm2.30$ mas,
respectively).  Correction to absolute parallax was performed using USNO
$BVI$ photometry for the individual reference stars along with a calibrated
$M_V$ versus $V-I$ relationship to derive a mean photometric parallax of
$0.85\pm0.15$ mas for the 29 star ensemble.  Together this then translates
to $\Pi_{\rm abs}= 51.6\pm0.8$ mas for LSPM J2322+7847.

\section{Activity and Ages}
The star HIP 115147 was identified with the bright X-ray source H2311+77, which led
\citet{pra} to suggest RS CVn-type activity. It was shown later on that the star is
not evolved, and that it is not a short-period spectroscopic binary, justifying the currently
accepted classification as a very young, ``naked" post-T Tauri dwarf \citep{nat, chu91, chu93}.
Being one of the most powerful X-ray emitters in the solar neighborhood \citep{vog} at $L_X=1.04
\cdot 10^{30}$ ergs s$^{-1}$ \citep{mak03}, the star has the same space velocity as the Local Young Stream
(or Local Association, \citet{mont}). This association limits its age to $\la 200$ Myr.
This stream includes isolated stars, groups and associations of diverse ages, some
as young as 1 Myr (e.g., in Lupus and Ophiuchus). Therefore, the assignment to this
stream by itself does not lend a more precise estimation of age. An X-ray luminosity of
$\log L_X \approx 30$ is typical of weak-lined TT stars in Taurus-Auriga-Perseus, but
significantly larger than that of classical TT stars; if anything, it points at an
age older than a few Myr. HIP 115147 is listed as variable star V368 Cep
\citep{kho}. The slight variability allowed \citet{kah} to determine the period of rotation
of this star, 2.74 d. The fast rotation is responsible for the high degree of chromospheric
and coronal activity. The primary star is identified as the extreme ultraviolet source
EUVE J$2319+790$ with strong detections at $100\AA$ band, as well as at 250 eV in X-rays
\citep{bow, lamp}. HIP 115147 is one of the 181 extreme-ultraviolet sources in the
ROSAT WFC all-sky Bright Source survey identified with late-type main-sequence stars
\citep{pou}, with high signal-to noise detections in both $60$-$140\AA$ and $110$-$200\AA$
passbands. An unusually high level of chromospheric activity of $\log R'_{\rm HK}=-4.074$ was
determined by \citet{gra}; a spectral type K2V is also specified in the latter paper as opposed
to G9V given in the Simbad database. Since the rate of rotation diminishes fairly quickly with age in single stars, so does X-ray luminosity, and open clusters older than $\alpha$ Persei (50 Myr)
are usually more quiescent than the youngest ones (IC 2602 and IC 2391). The high degree
of chromospheric and extreme ultraviolet activity suggests a very young age, possibly
less than 20 Myr. V368 Cep is more
powerful in X-rays than an average K-type Pleiades member by a factor of 10, indicating an age
less than 100 Myr. Finally, the equivalent width of \ion{Li}{1} at $W_{\rm Li}=0.218$ \AA~
is smaller than the upper limit for the Pleiades by $0.104$ \AA~ according to \citet{wich03},
which points at an age similar to the Pleiades or older. However, the lithium surface content 
is a poor estimator of age for young stars
because of the large intrinsic spread of this parameter in stars of the same mass and age.

\begin{figure}
\plotone{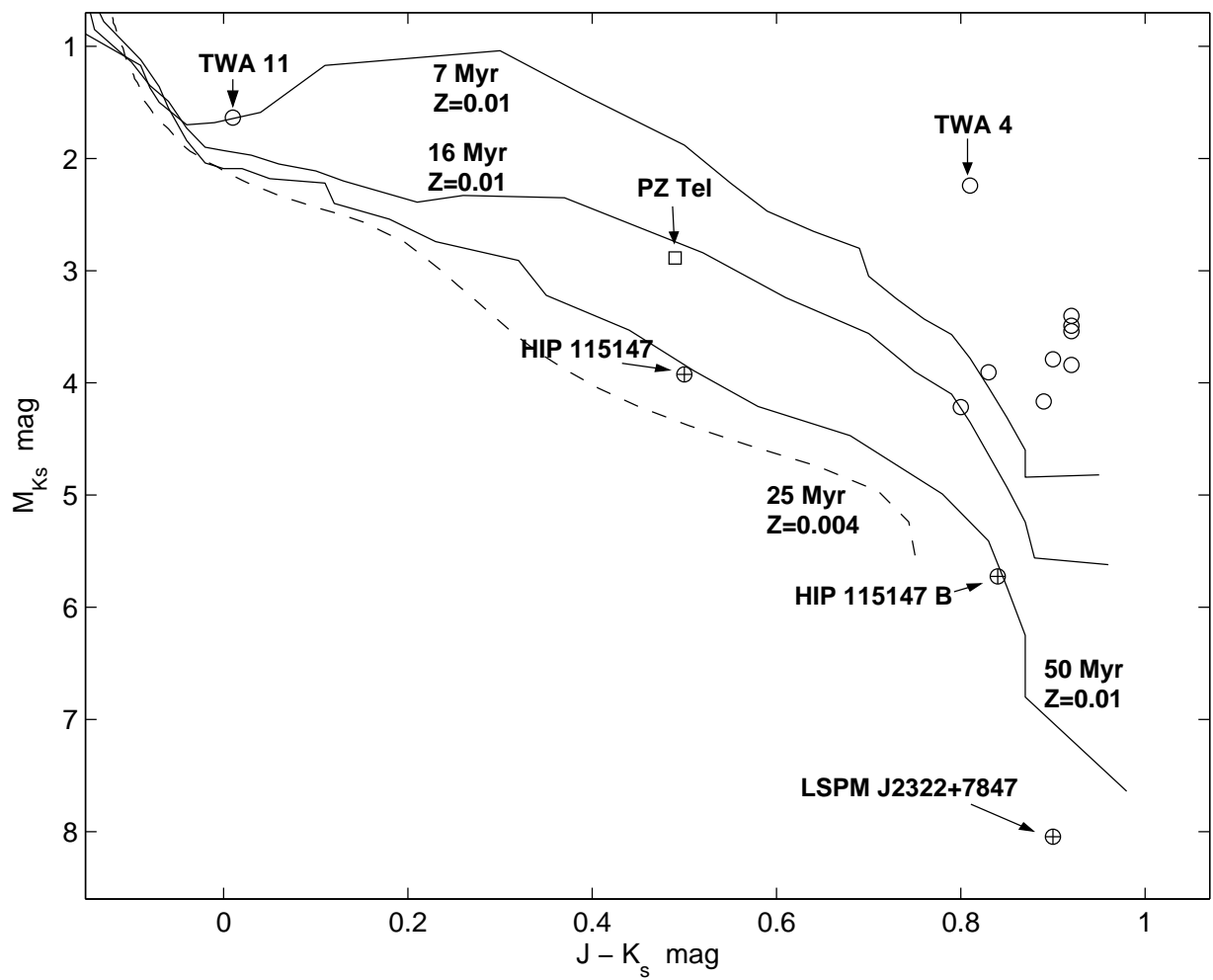}
\caption{Near-infrared HR diagram for HIP 115147 A and B components and for the
newly identified CPM companion LSPM J2322+7847 (circles with inscribed crosses).
For comparison, the young post-T Tauri star PZ Tel is shown with a square, as well as
members of the TWA group of pre-main-sequence stars, with open circles. 
Model isochrones from \citet{siess} for ages 7, 16 and 50 Myr and
metal abundance $Z=0.01$ are drawn with full lines and a 25 Myr isochrone with
$Z=0.004$ from \citep{piet} is indicated with a dashed line. \label{jk.fig}}
\end{figure}

\begin{figure}
\plotone{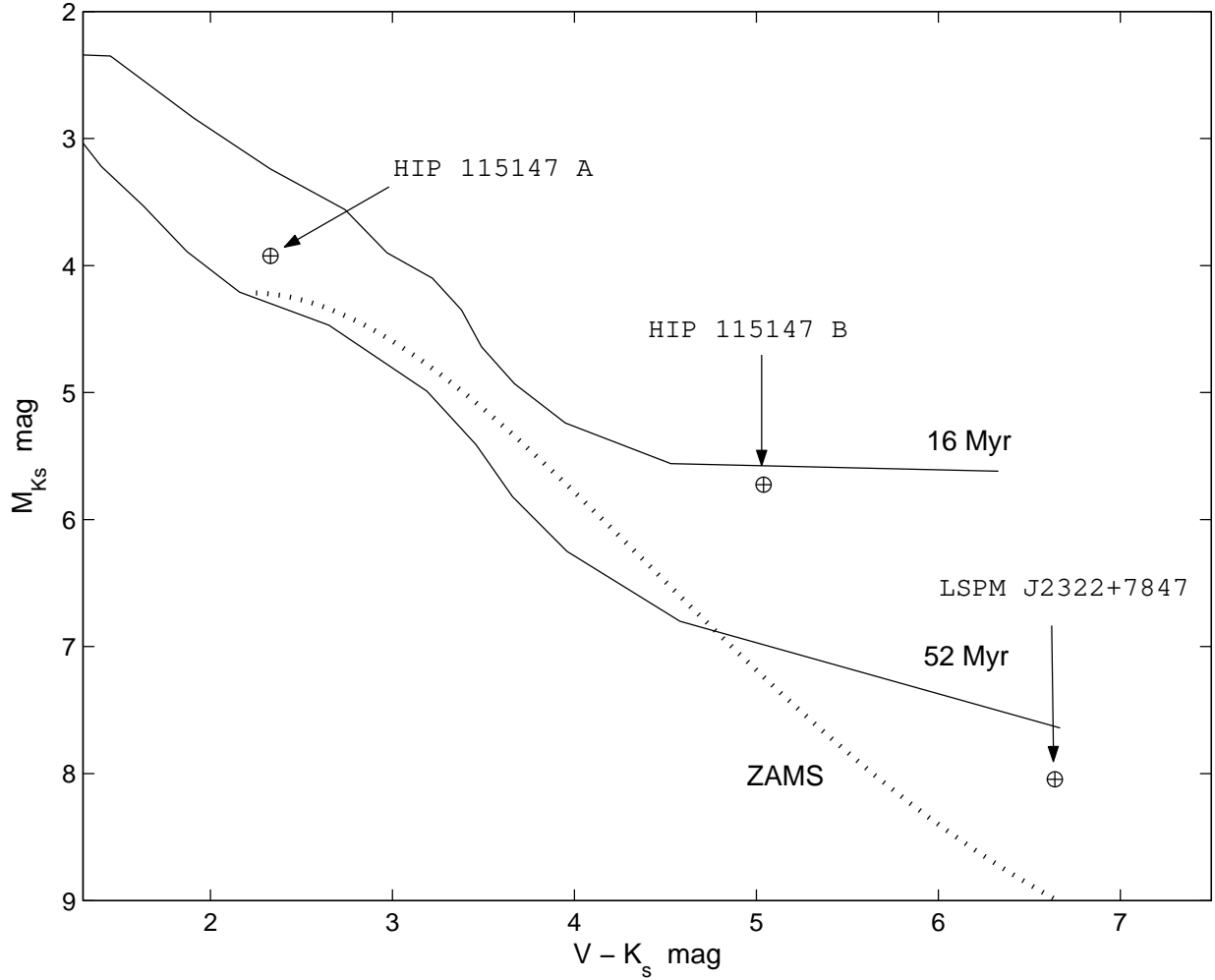}
\caption{$M_{Ks}$ versus $V-K_s$ HR diagram for HIP 115147 A, B and LSPM J2322+7847 components
indicated with circles with inscribed crosses.
Model isochrones from \citet{siess} for ages 16 and 52 Myr and
metal abundance $Z=0.01$ are drawn with full lines and the empirical main sequence
of field dwarfs from \citep{hen} is indicated with a thick dotted line. \label{vk.fig}}
\end{figure}

We are left with the most reliable and frequently used method of age estimation by
model isochrones. Using the photometric data from the literature and our own (Table~1),
we plot HR diagrams in the 2MASS $J$ and $K_s$, and $V$ passbands in Figs.~\ref{jk.fig}
and \ref{vk.fig}. For reference, the theoretical isochrones at $7$, $16$ and $52$ Myr
are drawn from \citet{siess}. According to the
data in \citet{nord} derived from Str{\"o}mgren {\it uvby$\beta$} photometry, the star HIP 115147 has a markedly subsolar metallicity at
[Fe/H]$=-0.64$. This determination has to be taken with caution, because the photometrically
derived metallicities are sensitive to the color index $m_1$ (B. Nordstr{\"o}m 2007, priv. comm.),
which may be abnormally high for chromospherically active stars \citep{mor, fav}.
A wider range of metal abundances is supplied in the evolution
models of \citet{piet}, and we present also a 25 Myr isochrone for $Z=0.004$ from their
models in Fig.~\ref{jk.fig} with a dashed line. We find that the Siess et al. 50 Myr
isochrone ($Z=0.01$) provides the best fit for all three alleged comoving components (marked with crossed circles), including the secondary M dwarf HIP 115147 B, separated by
$\sim 11 \arcsec$ from the primary \citep{low}. This pair is listed in the WDS \citep{mas} as
WDS 23194$+7900$ (LDS 2035, discovered by Luyten in 1969), at separation $10.8\arcsec$ and
position angle $216\degr$. The relative position did not change significantly between the first
AC2000.2 measures taken in 1901 and the latest epoch 2000 in 2MASS (W. Hartkopf \& G. Wycoff 2007,
priv. comm.), confirming that the inner pair is physical. The small deviation of the faintest component 
LSPM J2322+7847 from the Siess 50 Myr isochrone appears to arise mostly from a sudden twist of the
isochrone at the latest data point corresponding to mass $0.1 M_{\sun}$. We are not in a position
to discuss if this twist has a certain physical meaning, but the overall match is good,
and the estimated mass of the star is roughly $0.1 M_{\sun}$. 

The younger isochrones for 16 and 7 Myr from \citet{siess} lie significantly higher in
Fig.~\ref{jk.fig} than the matching 50 Myr isochrone. The 25 Myr isochrone
from \citet{piet} at a lower metallicity provides a poor fit to
our stars, predicting
much bluer colors, but these models may be strongly biased for young ages as was found
by \citet{mak} for solar-type stars in the Alpha Persei open
cluster  at 52 Myr. The members of the very young TWA association are shown
in Fig.~\ref{jk.fig} with open circles. These stars are certainly much younger than
HIP 115147 and its companions. They appear to be brighter and redder than the 7 Myr isochrone corresponding to
the probable age of this group \citep[see also][]{mak05}, but this deviation may be the
result of the near-infrared $K$ and $J$ excess commonly observed in classical T Tauri
stars and attributed to a hot inner rim in their dusty accretion disks \citep{cie}. Since our
stars under investigation have no accretion disks and
their metal abundance may be lower, we need to find a young star with similar parameters.
It was pointed out by \citet{mak07} that a few young stars currently in the solar neighborhood,
that traveled from the vicinity of the Ophiuchus star forming region,
share the moderately poor metal abundance of HIP 115147 as determined from the {\it uvby$\beta$}
photometry, probably affected by the high degree of chromospheric activity. One of these stars is the
extremely active dwarf PZ Tel, a likely member of the $\beta$ Pic associations estimated to
be 12-20 Myr old. This star is indicated with an open square in Fig.~\ref{jk.fig}. Its
position matches the 16 Myr, $Z=0.01$ isochrone quite well.
PZ Tel is significantly brighter
than HIP 115147 in $K_s$ having approximately the same color; thus, the
latter star and its companions are probably older than 16 Myr. It is verified in Fig.
\ref{vk.fig} that this difference in $M_{Ks}$ between the two stars is not originating in a
$K_s$-band excess, since both HIP 115147 and LSPM J2322+7847 match best the 52 Myr, $Z=0.01$
isochrone in a $M_{Ks}$ versus $V-K_s$ diagram as well. The latter star is brighter in $K_s$ than
the empirically determined main sequence for field dwarfs from \citet{hen} by 1 mag. If
this star were an unrelated old M6 dwarf, its distance would be only 12 pc. Our trigonometric parallax
(Table 1) yields a distance $19.4\pm0.3$ pc.

\section{Concluding Remarks}
A number of stars with outstanding signs of activity and young age scattered in the solar
vicinity can be traced back to their places of origin in, or close to, the OB associations
in the Sco-Cen complex \citep{wis}.
The star HIP 115147 has come from the vicinity of the molecular cloud LDN 1709 in the
Ophiuchus star forming region, flanking the Sco-Cen complex at $\ell\approx 0$ 
\citep{mak07}. The closest approach to the estimated center of that association
is 10.7 pc, while the separation today is 170 pc. The closest approach 16 Myr ago saw
the star flying by at a relative velocity of 10.8 \kms. If HIP 115147 was born in the
Ophiuchus association, its probable age should be 16 Myr, considerably younger than the
previous isochrone analysis and the Lithium abundance suggest. Because of the high
departing velocity, it is more likely that
this close fly-by is a chance occurrence. 

Finally, we note that the difference in proper motion between HIP 115147 and LSPM J2322+7847,
which appears to be statistically significant, may be caused by a systematic error of 5 to 10 \masyr~ in
the NOMAD proper motions for very faint stars. The formal errors specified in NOMAD may be
strongly underestimated for this optically dim star, whose position at the mean epoch 1979
is based on Schmidt plates. The proper motion difference, if it is true, implies
that the two stars are not gravitationally bound as a truly multiple system. This is
commonly observed among very young stars in the solar neighborhood; for example, the
common proper motion pair AT Mic and AU Mic (approximately 10 Myr old) are separated by at least 0.23 pc and
have proper motions different by $\simeq 7$\%, but their physical association is not in doubt.
At the
suggested age (20-50 Myr), the HIP 115147 companions are likely to stay close to each other as a moving
group. More accurate absolute proper motions for both companions and
spectroscopic radial velocities are required to clarify the dynamical status of the system. 
The currently available evidence
is consistent with LSPM J2322+7847 being one of the youngest, later-type stars in the near solar neighborhood.

\acknowledgments
We thank T. Tilleman and C. Dahn for taking some of the
parallax images. The research described in this paper was in part carried out at the Jet Propulsion 
Laboratory, California Institute of Technology, under a contract with the National 
Aeronautics and Space Administration. This research has made use of the SIMBAD database,
operated at CDS, Strasbourg, France; and data products from the 2MASS, which is a joint project
of the University of Massachusetts and the Infrared Processing and Analysis Center, California
Technology Institute, funded by NASA and the NSF.

\clearpage 

\begin{deluxetable}{llll}
\tabletypesize{\scriptsize}
\tablecaption{Astrometry and photometry of HIP 115147 A, B and LSPM J2322+7847 \label{tab}}
\tablewidth{0pt}
\tablehead{
\colhead{Name} & \colhead{HIP 115147 A} & \colhead{HIP 115147 B} &
\colhead{LSPM J2322+7847}}
\startdata

RA J2000  \dotfill & 23 19 26.632 & 23 19 24.53 & 23 22 53.873  \\
Dec. J2000  \dotfill & +79 00 12.67  & +79 00 03.8 &  +78 47 38.81 \\
$\Pi$  \dotfill& $50.7\pm0.6$ &  & $51.6\pm0.8$ \\
Proper motion \dotfill& $(201,72)\pm(1,1)$ & $(203,72)$ & $(210,64)\pm(2,3)$ \\
$B$ \dotfill&8.60 & 13.75 & 17.99 \\
$V$ \dotfill& 7.73 &12.24& 16.16 \\
$R$ \dotfill& 7.17 & 11.04& 14.61 \\
$I$ \dotfill& 6.76 & 9.54 & 12.54 \\
$J$ \dotfill& 5.90 & 8.04 & 10.42 \\
$H$ \dotfill& 5.51 & 7.39 & 9.84 \\
$K_s$ \dotfill& 5.40 & 7.20 & 9.52 \\

\enddata
\tablecomments{Right Ascension is in $(h,m,s)$, Declination is in $(\degr,\arcmin,\arcsec)$,  proper motion in \masyr, parallax in mas. Parallax for HIP 115147 A is from Hipparcos, parallax for
LSPM J2322+7847 is our observation. Proper motion for HIP 115147 A is from Hipparcos, 
for HIP 115147 B from \citep{goch}, and for LSPM J2322+7847 from NOMAD. $BVRI$ magnitudes are our measurements. 
$JHK_s$ magnitudes are from 2MASS. 
The relative position of the faint companion at J2000 is separation $962.56\arcsec$,
position angle $141.1\degr$.}
\end{deluxetable}

\clearpage

\clearpage

\end{document}